\documentclass[preprint,preprintnumbers, prd, floatfix, superscriptaddress,nofootinbib] {revtex4-1}
\usepackage{epsfig}
\usepackage{subfigure}
\usepackage{dcolumn}
\usepackage{bm}
\usepackage[usenames ,dvipsnames]{xcolor}
\usepackage{slashed}
\usepackage{graphicx,color}
\begin{document}
\title{Hidden charm ${\cal P}_{cs}(4338)^0$ production in baryonic 
$B^-\to J/\psi \Lambda\bar p$ decay}

\author{Yu-Kuo Hsiao}
\email{yukuohsiao@gmail.com}
\affiliation{School of Physics and Information Engineering, 
Shanxi Normal University, Taiyuan 030031, China}

\author{Shu-Ting Cai}
\email{18734581917@163.com}
\affiliation{School of Physics and Information Engineering, 
Shanxi Normal University, Taiyuan 030031, China}

\author{Yan-Li Wang}
\email{ylwang0726@163.com}
\affiliation{School of Physics and Information Engineering, 
Shanxi Normal University, Taiyuan 030031, China}

\date{\today}

\begin{abstract}
We investigate the resonant baryonic $B$ decay
$B^-\to {\cal P}_{cs}^0\bar p,{\cal P}_{cs}^0\to J/\psi \Lambda$,
where ${\cal P}_{cs}^0\equiv {\cal P}_{cs}(4338)^0$ 
is identified as a hidden charm pentaquark candidate with strangeness.
By interpreting ${\cal P}_{cs}^0$ as the $\Xi_c\bar D$ molecule
that strongly decays into $J/\psi\Lambda$ and $\eta_c\Lambda$, 
we discover a dominant triangle rescattering effect for $B^-\to {\cal P}_{cs}^0\bar p$, 
initiated by $B^-\to J/\psi K^-$. 
Through the exchange of a $\bar \Lambda$ anti-baryon, 
$J/\psi$ and $K^-$ undergo rescattering, transforming into ${\cal P}_{cs}^0$ and $\bar p$, 
respectively. Based on this rescattering mechanism,
we calculate ${\cal B}(B^-\to {\cal P}_{cs}^0\bar p,{\cal P}_{cs}^0\to J/\psi \Lambda)
=(1.4^{+2.1}_{-1.1})\times10^{-6}$, which is consistent with the measured data.
\end{abstract}

\maketitle
\section{introduction}
In three-body anti-triplet $b$-baryon decays,
${\bf B}_b\to J/\psi {\bf B}M$ plays a key role in exploring 
the hidden charm pentaquark states
${\cal P}_c(c\bar c uud)$ and ${\cal P}_{cs}(c\bar c uds)$~\cite{Chen:2022asf,Johnson:2024omq},
where ${\bf B}(M)$ denotes an octet baryon (meson). 
For instance, ${\cal P}_c(4380,4450)^+$ 
were observed as the first  pentaquark candidates 
in $\Lambda_b\to J/\psi p K^-$~\cite{LHCb:2015yax}. Later, 
LHCb reported the observation of ${\cal P}_c(4312)^+$~\cite{LHCb:2019kea}.
In the same report, ${\cal P}_c(4450)^+$ due to higher resolution was split into 
${\cal P}_c(4440,4457)^+$, while ${\cal P}_c(4380)^+$ became an uncertain state.
Additionally, ${\cal P}_{cs}(4459)^0$ was newly found 
with strangeness in $\Xi_b^-\to J/\psi \Lambda K^-$~\cite{LHCb:2020jpq}.

In three-body baryonic $B$ decays, $B\to J/\psi {\bf B\bar B'}$ is proposed to
provide an equally important test bed for studying ${\cal P}_{c(s)}$~\cite{Brodsky:1997yr}.
Indeed, ${\cal P}_c^\pm\equiv{\cal P}_c(4337)^\pm$ was observed 
in $\bar B^0_s\to J/\psi p\bar p$~\cite{LHCb:2021chn}. Moreover,
${\cal P}_{cs}^0\equiv {\cal P}_{cs}(4338)^0$
has been observed with $(m,\Gamma)=(4338.2\pm 0.7\pm 0.4,7.0\pm 1.2\pm 1.3)$~MeV
in $B^-\to J/\psi \Lambda\bar p$
and has been assigned the quantum numbers $J^P=1/2^-$~\cite{LHCb:2022ogu}.
Since the resonant branching fraction
${\cal B}(\bar B^0_s\to \bar p({\cal P}_c^+\to)J/\psi p+p({\cal P}_c^-\to) J/\psi \bar p)$ and 
${\cal B}(B^-\to \bar p({\cal P}_{cs}^0\to)J/\psi\Lambda)$ 
are found to be $(22.0^{+8.5}_{-4.0}\pm 8.6)\%$ and $(12.5\pm 0.7\pm 1.9)\%$ 
of the total branching fractions 
${\cal B}(\bar B^0_s\to J/\psi p\bar p)=(3.6\pm 0.4)\times 10^{-6}$ and 
${\cal B}(B^-\to J/\psi \Lambda\bar p)=(14.6\pm 1.2)\times 10^{-6}$~\cite{pdg}, respectively,
this leads to
\begin{eqnarray}\label{data1}
&&
{\cal B}(\bar B^0_s\to \bar p({\cal P}_c^+\to)J/\psi p+p({\cal P}_c^-\to) J/\psi \bar p)
=(7.9\pm 4.4)\times 10^{-7}\,,\nonumber\\
&&
{\cal B}(B^-\to \bar p({\cal P}_{cs}^0\to)J/\psi\Lambda)
=(1.8\pm 0.3)\times 10^{-6}\,.
\end{eqnarray}
It is perplexing that there are no mutual discoveries of ${\cal P}_{c(s)}$ states 
between ${\bf B}_b\to J/\psi {\bf B}M$ and 
$B\to J/\psi {\bf B\bar B'}$~\cite{Nakamura:2021dix,Yan:2021nio,Wu:2024lud}. 
To resolve this confusion, $B\to J/\psi {\bf B\bar B'}$ 
should be further explored~\cite{Hsiao:2009mk,Chen:2008sw}, 
particularly for the production of ${\cal P}_{c(s)}$ in 
$B\to {\cal P}_{c(s)}\bar{\bf B}'$ with the subsequent strong decay of
${\cal P}_{c(s)}\to{\bf B}J/\psi$.

Due to the short-distance picture, $B\to{\bf B \bar B'}$ is suppressed 
by hard QCD effect and helicity conservation~\cite{Hou:2000bz,Suzuki:2006nn,
Hou:2019uxa,Chua:2022wmr,Hsiao:2019wyd,Hsiao:2014zza,Huang:2021qld,Hsiao:2023mud},
as reflected in the measurements of 
${\cal B}(\bar B^0_s\to p\bar p)<4.4\times 10^{-9}$~\cite{LHCb:2022oyl}
and ${\cal B}(B^-\to \Lambda \bar p)\simeq 2\times 10^{-7}$~\cite{pdg}.
One can view $B\to {\cal P}_{c(s)}\bar{\bf B}'$ as an exotic edition of 
two-body baryonic $B$ decay, where ${\cal P}_{c(s)}$ acquires an additional
$c\bar c$ pair from the sea quarks or intrinsic charm in $B$ meson~\cite{Brodsky:1980pb,Brodsky:1981se,Brodsky:1997fj,Chang:2001iy,
Brodsky:2001yt,Mikhasenko:2012km,Zhou:2017bhq,Hsiao:2015nna}.
Consequently, $B\to{\cal P}_{c(s)} \bar{\bf B}'$ should be as suppressed as $B\to{\bf B\bar B}'$.
Since ${\cal B}(\bar B^0_s\to {\cal P}_c^+\bar p)$ is 10 to 100 times larger than
the experimental upper limit of ${\cal B}(\bar B^0_s\to p\bar p)$, and 
${\cal B}(B^-\to {\cal P}_{cs}^0 \bar p)$ is 10 times larger than ${\cal B}(B^-\to \Lambda \bar p)$,
it is reasonable to infer that 
$B\to {\cal P}_{c(s)}{\bf\bar B'}$ actually proceeds via the long-distance 
final state interaction (FSI), thereby avoiding the short-distance suppression.
Thus, we propose calculating the FSI triangle rescattering processes for ${\cal P}_{c(s)}$ production
in $B\to J/\psi{\bf B\bar B}'$ as a suitable mechanism
for interpreting the resonant branching fractions in Eq.~(\ref{data1}).

\section{Formalism}
%
\begin{figure}[t!]
\includegraphics[width=2.5in]{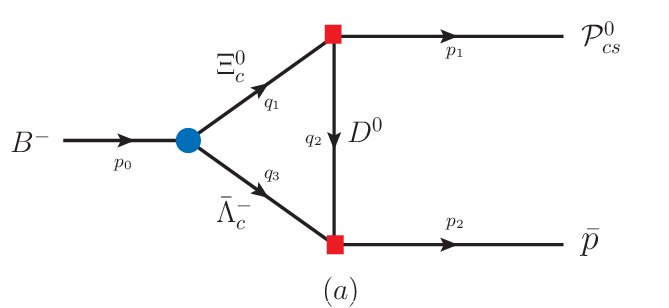}
\includegraphics[width=2.5in]{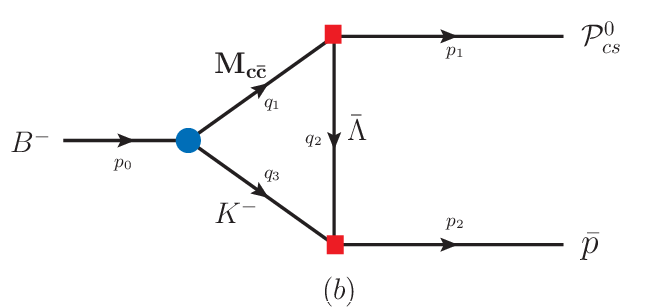}
\includegraphics[width=2.5in]{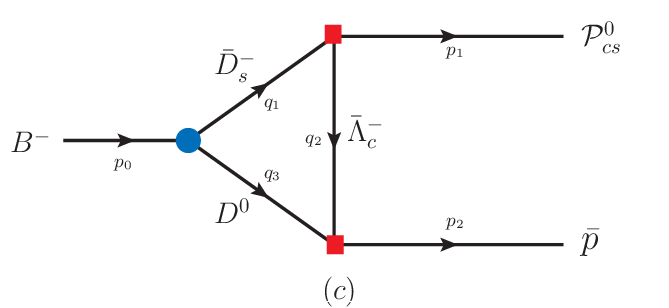}
\caption{Triangle rescattering diagrams of $B^-\to {\cal P}_{cs}^0\bar p$,
where 
(a) depicts $B^-\to \Xi_c^0\bar \Lambda_c^-\to {\cal P}_{cs}^0\bar p$ via $D^0$ meson exchange,
(b) depicts $B^-\to M_{c\bar c} K^-\to {\cal P}_{cs}^0\bar p$, with $M_{c\bar c}=\eta_c$ or $J/\psi$, 
via $\bar \Lambda$ baryon exchange, and (c) depicts $B^-\to D^0\bar D_s^-\to{\cal P}_{cs}^0\bar p$ 
via $\bar \Lambda_c^-$ baryon exchange.}\label{fig1}
\end{figure}
%
 
To establish the rescattering mechanism 
for the hidden-charm pentaquark production in baryonic $B$ decays, 
we prefer to use $B^-\to \bar p({\cal P}_{cs}^0\to)J/\psi\Lambda$ as an illustrative example.
This is due to the fact that the ${\cal P}_{cs}^0$ strong decays,
studied in Refs.~\cite{Azizi:2023iym,Ortega:2022uyu,Wang:2024rsm},
can provide strong coupling constants required for the triangle loop.
On the other hand, information regarding the strong decays of
${\cal P}_c^\pm$ is nearly unavailable, which means that
$\bar B^0_s\to \bar p({\cal P}_c^+\to)J/\psi p$ and 
$\bar B^0_s\to p({\cal P}_c^-\to) J/\psi \bar p$ are currently beyond the scope 
of our investigation.

As depicted in Figs.~\ref{fig1}(a, b) and (c), the FSI triangle rescattering effects 
are responsible for the production of ${\cal P}_{cs}^0$ in $B^-\to {\cal P}_{cs}^0 \bar p$.
It is interesting to note that the rescattering diagrams similar to
Figs.~\ref{fig1}(a, c) have once been proposed 
to induce triangle singularity~\cite{Burns:2022uha}.

In Fig.~\ref{fig1}a, $B^-\to {\cal P}_{cs}^0\bar p$ starts with $B^-\to \Xi_c^0\bar \Lambda_c^-$,
where $\Xi_c^0$ and $\bar \Lambda_c^-$, by exchanging $D^0$, 
undergo rescattering, transforming into ${\cal P}_{cs}^0$ and $\bar p$, respectively. 
This rescattering  incorporates the amplitudes of 
$B^-\to \Xi_c^0\bar \Lambda_c^-$, $\Xi_c^0 \to{\cal P}_{cs}^0 D^0$, 
and $\bar\Lambda_c^-\to \bar p \bar D^0$,
as given by
\begin{eqnarray}\label{fig1a}
&&
{\cal M}_{a1}\equiv {\cal M}(B^-\to \Xi_c^0\bar \Lambda_c^-)
=\frac{G_F}{\sqrt 2}V_{cb}V_{cs}^* \bar{u}_{\Xi_c}(F-G\gamma_5)v_{\bar\Lambda_c}\,,\nonumber\\
&&
{\cal M}_{a2}\equiv {\cal M}(\Xi_c^0 \to{\cal P}_{cs}^0 D^0)
=f_{\Xi_c D}\bar u_{{\cal P}_{cs}} u_{\Xi_c}\,,\nonumber\\
&&
{\cal M}_{a3}\equiv {\cal M}(\bar\Lambda_c^-\to \bar p \bar D^0)=
g_{pD}\bar v_{\bar\Lambda_c}\gamma_5 v_{\bar p}\,.
\end{eqnarray}
Since $B^-\to \Xi_c^0\bar \Lambda_c^-$ has been observed
with ${\cal B}\sim 10^{-3}$~\cite{pdg}, and studied 
with the non-factorizable contribution~\cite{Hsiao:2023mud},
the paramters $F$ and $G$ in ${\cal M}_{a1}$ are extractable. 
For the strong coupling constants in the triangle loop,
$f_{\Xi_c D}$ and $g_{pD}$ can be found 
in~\cite{Azizi:2023iym,Ortega:2022uyu,Wang:2024rsm} 
and~\cite{Duan:2023dky}, respectively. Therefore, 
the required information is sufficient for our calculation.

In Fig.~\ref{fig1}b, $B^-\to M_{c\bar c} K^-$ with  $M_{c\bar c}\equiv (J/\psi,\eta_c)$ 
proceeds as the starting weak decay. Through the $\bar\Lambda$ anti-baryon exchange, 
$M_{c\bar c}$ and $K^-$ rescatter with each other, 
transforming into ${\cal P}_{cs}^0$ and $\bar p$, respectively.
This rescattering effect requires the amplitudes of $B^-\to M_{c\bar c} K^-$, 
$ M_{c\bar c}\to{\cal P}_{cs}^0\bar\Lambda$, and $\bar\Lambda\to \bar pK^+$, 
written as
\begin{eqnarray}\label{fig1b}
&&
{\cal M}_{b1}\equiv {\cal M}(B^-\to M_{c\bar c} K^-)
=\frac{G_F}{\sqrt 2}V_{cb}V_{cs}^* a_2^{M_{c\bar c}}
\langle M_{c\bar c}|(\bar c c)|0\rangle \langle K^{-}|(\bar s b)|B^-\rangle\,,\nonumber\\
&&
{\cal M}_{b2}\equiv {\cal M}(\eta_c\to{\cal P}_{cs}^0\bar\Lambda)
=f_{\Lambda\eta_c}\bar u_{{\cal P}_{cs}} v_{\bar\Lambda}\,,\nonumber\\
&&
{\cal M}'_{b2}\equiv {\cal M}(J/\psi \to {\cal P}_{cs}^0\bar \Lambda)
=\epsilon_\mu \bar u_{{\cal P}_{cs}}\bigg[g_{\Lambda J}\gamma^\mu
-i\frac{h_{\Lambda J}}{(m_{{\cal P}_{cs}}+m_\Lambda)}\sigma^{\mu\nu}q_\nu\bigg]\gamma_5 v_{\bar\Lambda}\,,\nonumber\\
&&
{\cal M}_{b3}\equiv {\cal M}(\bar\Lambda\to \bar p K^+)
=g_{pK}\bar v_{\bar\Lambda} \gamma_5 v_{\bar p}\,,
\end{eqnarray}
where $(\bar q_i q_j)\equiv \bar q_i\gamma_\mu(1-\gamma_5) q_j$.
Due to the factorization approach~\cite{Gourdin:1994xx,Cheng:1998kd},
${\cal M}_{b1}$ is decomposed as $a_2^{M_{c\bar c}}$,
the matrix elements of the $B$ to $K$ transition, and the matrix elements of 
the vacuum (0) to $M_{c\bar c}$ production, which has been used to explain the data~\cite{pdg}.
The association with ${\cal P}_{cs}^0\to \eta_c\Lambda$ and 
${\cal P}_{cs}^0\to J/\psi\Lambda$~\cite{Azizi:2023iym,Ortega:2022uyu,Wang:2024rsm}
can lead to the extraction of
$f_{\Lambda\eta_c}$ and $(g_{\Lambda J},h_{\Lambda J})$, respectively.
For $g_{pK}$, it is obtained from ${\bf B}\to {\bf B}'M$ 
under $SU(3)_f$ symmetry~\cite{Yang:2018idi}.
Clearly, the rescattering effect shown in Fig.~\ref{fig1}(b) can also be computed,
with the aforementioned couplings incorporated into the integration of the triangle loop.

In Fig.~\ref{fig1}(c), $B^-\to {\cal P}_{cs}^0\bar p$ initially proceeds through $B^-\to D^0\bar D_s^-$.
In the triangle loop, $\bar D_s^-$ decays into ${\cal P}_{cs}^0$ and $\bar \Lambda_c^-$,
while $D^0$, interacting with $\bar \Lambda_c^-$, is converted to $\bar p$.
The required amplitudes are expressed as
\begin{eqnarray}\label{fig1c}
&&
{\cal M}_{c1}\equiv {\cal M}(B^-\to D^0\bar D_s^-)
=\frac{G_F}{\sqrt 2}V_{cb}V_{cs}^{*} a_1^{D} 
\langle \bar{D}_s^-|(\bar s c)|0\rangle \langle  D^0|(\bar c b)|B^-\rangle\,,\nonumber\\
&&
{\cal M}_{c2}\equiv {\cal M}(\bar D_s^-\to{\cal P}_{cs}^0\bar\Lambda_c^-)
=f_{\Lambda_c D_s}\bar u_{{\cal P}_{cs}}v_{\bar\Lambda_c}\,,\nonumber\\
&&
{\cal M}_{c3}={\cal M}_{a3}\,,
\end{eqnarray}
where ${\cal M}_{c1}$ is also derived with the factorization~\cite{Cheng:1998kd}.
In addition to $f_{\Lambda_c D_s}$, which is extracted from 
$\Gamma({\cal P}_{cs}^0\to \Lambda_c^+\bar D_s^-)$~\cite{Azizi:2023iym,Ortega:2022uyu,Wang:2024rsm},
we will be able to integrate the triangle loop depicted in Fig.~\ref{fig1}c.

For ${\cal M}(B^-\to M_{c\bar c} K^-)$ in Eq.~(\ref{fig1b}) and 
${\cal M}(B^-\to D^0\bar D_s^-)$ in Eq.~(\ref{fig1c}),
the matrix elements of the vacuum (0) to meson production and the $B$ to $P_q$ transition
can be parameterized as~\cite{Gourdin:1994xx,Cheng:1998kd}
\begin{eqnarray}\label{ffs}
&&
\langle J/\psi|(\bar cc)|0\rangle=m_{J/\psi} f_{J/\psi} \epsilon_\mu^*\,,\nonumber\\
&&
\langle P_{q\bar c}|(\bar q c)|0\rangle=if_{P_{q\bar c}}p_\mu\,,\nonumber\\
&&
\langle P_q|(\bar q b)|B\rangle=
F_1^{P_q}\bigg[(p_B+p_{P_q})^\mu-\frac{m_B^2-m_{P_q}^2}{p^2}p^\mu\bigg]+
F_0^{P_q}\frac{m_B^2-m_{M_{P_q}}^2}{p^2}p^\mu\,,
\end{eqnarray}
where $p_\mu=(p_B-p_{P_q})_\mu$, 
$q=c$ is for $(P_{q\bar c},P_q)=(\eta_c,D^0)$, and
$q=s$ is for $(P_{q\bar c},P_q)=(\bar D_s^-,K^-)$, 
with $f_{J/\psi}$ and $f_{P_{q\bar c}}$ being the decay constants. 
In Eq.~(\ref{ffs}), the momentum-dependent form factors
$F_{1,0}^{P_q}$ are represented in a three-parameter form~\cite{MS}:
\begin{eqnarray}\label{form2}
&&
F_1^{P_q}(p^2)=
\frac{F_1^{P_q}}{(1-\frac{p^2}{M_A^2})
(1-\sigma_{11}^{P_q} \frac{p^2}{M_A^2}+\sigma_{12}^{P_q}\frac{p^4}{M_A^4})}\,,\;\nonumber\\
&&
F_0^{P_q}(p^2)=
\frac{F_0^{P_q}}{1-\sigma_{01}^{P_q}\frac{ p^2}{M_A^2}+\sigma_{02}^{P_q}\frac{ p^4}{M_A^4}}\,,
\end{eqnarray}
where $F_{1,0}^{P_q}\equiv F_{1,0}^{P_q}(0)$ are at $p^2=0$, and
$\sigma_{j1,j2}^{P_q}$ ($j=0,1$) are parameters along with the pole mass $M_A$.

Utilizing Eqs.~(\ref{fig1a}), (\ref{fig1b}), and (\ref{fig1c}), 
we can express the triangle rescattering amplitudes of Fig.~\ref{fig1}(a, b) and (c) 
as follows~\cite{Hsiao:2019ait,Yu:2020vlt,Yu:2021euw,Hsiao:2021tyq}:
\begin{eqnarray}\label{amp_abc}
{\cal M}_i^{(\prime)}=\int \frac{d^4{q}_{1}}{(2\pi)^{4}}
\frac{{\cal M}_{i1}{\cal M}_{i2}^{(\prime)}{\cal M}_{i3}F_i(q_{2}^2)}
{(q_{1}^{2}-{m_1}^{2})[(q_1-p_1)^{2}-m_{2}^{2}][(q_1-p_0)^{2}-m_{3}^{2}]}\,,
\end{eqnarray}
where $i$=(a, b, c). Here, we define
${\cal M}_a={\cal M}(B^-\to \Xi_c^0\bar \Lambda_c^-\to {\cal P}_{cs}\bar p)$,
${\cal M}_b={\cal M}(B^-\to \eta_c K^-\to {\cal P}_{cs}\bar p)$, 
${\cal M}_b^\prime={\cal M}(B^-\to J/\psi K^-\to {\cal P}_{cs}\bar p)$, 
and
${\cal M}_c={\cal M}(B^-\to D^0\bar D_s^-\to{\cal P}_{cs}\bar p)$. 
In the above equation, $(p_0^\mu,p_1^\mu,p_2^\mu)$ and $(q_1^\mu,q_2^\mu,q_3^\mu)$ 
are the four-momentum energy flows, assigned in the triangle diagrams of Figs.~\ref{fig1}(a, b, c).
To avoid the divergence of the integration~\cite{Du:2021zdg}, we introduce
the form factor
$F_i(q_{2}^2)\equiv(\Lambda_{\rm cut}^{2}-m^{2}_2)/(\Lambda_{\rm cut}^{2}-q^{2}_{2})$,
where $\Lambda_{\rm cut}$ is a universal cutoff parameter. 
The choice of $\Lambda_{\rm cut}$ is based on the fact that 
the three rescattering amplitudes pertain to the same decay process,
thus requiring a consistent cutoff scale for all.

The ${\cal P}_{cs}$ state is observed as a resonance 
in the three-body decay $B^-\to {\cal P}_{cs}\bar p,{\cal P}_{cs}\to J/\psi\Lambda$~\cite{LHCb:2022ogu}.
Accordingly, we present the resonant amplitude as~\cite{Hsiao:2023qtk,Tsai:2021ota}
\begin{eqnarray}\label{3b_amp}
&&
{\cal M}(B^-\to {\cal P}_{cs}\bar p,{\cal P}_{cs}\to J/\psi\Lambda)\nonumber\\
&&={\cal M}'_{b2}({\cal P}_{cs}\to J/\psi\Lambda){\cal D}^{-1}{\cal M}_i^{(\prime)}(B^-\to{\cal P}_{cs}\bar p)\,,
\end{eqnarray}
where ${\cal M}'_{b2}({\cal P}_{cs}\to J/\psi\Lambda)$ is defined in Eq.~(\ref{fig1b}) and 
${\cal M}_i^{(\prime)}(B^-\to{\cal P}_{cs}\bar p)$ is calculated in Eq.~(\ref{amp_abc}).
Additionally, ${\cal D}\equiv(p_{J/\psi}+p_{\Lambda})^2-i m_{{\cal P}_{cs}} \Gamma_{{\cal P}_{cs}}$
corresponds to the Breit-Wigner form for the resonant ${\cal P}_{cs}$ state~\cite{pdg}.
On the other hand, 
the non-resonant amplitude of $B^-\to J/\psi\Lambda\bar p$ can be found in
Refs.~\cite{Hsiao:2014tda,Hsiao:2013dta,Cheng:2012fq,Chen:2008sw}, 
where the calculations are based on the factorization approach~\cite{Geng:2005fh,Geng:2007cw}.
To integrate over the phase space in the (non-)resonant three-body decay,
we employ the equation for the decay width as~\cite{pdg}
\begin{eqnarray}\label{gamma1}
\Gamma=\int_{m_{J\Lambda}^2}\int_{m_{\Lambda\bar p}^2}
\frac{1}{(2\pi)^3}\frac{|\bar {\cal M}|^2}{32M^3_B}dm_{J\Lambda}^2 dm_{\Lambda\bar p}^2\,,
\end{eqnarray}
with $m_{J\Lambda}=p_{J/\psi}+p_{\Lambda}$ and $m_{\Lambda\bar p}=p_{\Lambda}+p_{\bar p}$,
where $|\bar {\cal M}|^2$ represents the squared amplitude. 
The bar notation indicates the summation over the baryon spins.

\section{Numerical analysis}
%
\begin{table}[b]
\caption{Information on ${\cal P}_{cs}^0$ decays, 
where the decay width $\Gamma$ is in units of MeV. In S1,
the value with the star notation is due to our calculation with the decay width.}\label{tab1}
{\tiny
\begin{tabular}
{|l|c|c|c|} 
\hline
&S1 (Ortega)~\cite{Ortega:2022uyu}
&S2 (Azizi)~\cite{Azizi:2023iym}
&S3 (Wang)~\cite{Wang:2024rsm}\\
\hline
$\Gamma_{\Xi_c D},f_{\Xi_c D}$
&1.1, 0.29$^*$
&---
&---
\\ 
$\Gamma_{\eta_c\Lambda},f_{\eta_c\Lambda}$
&1.2, 0.15$^*$
&$3.18\pm 0.74$, $0.11\pm 0.02$ 
&$0.95\pm 0.05$, $0.184\pm 0.048$ 
\\	
$\Gamma_{J\Lambda},g_{J\Lambda},h_{J\Lambda}$
&0.6, 0.23$^*$, 0.23$^*$
&$7.22\pm 1.78$, $(-4.71\pm 0.52)\times 10^{-5}$, $0.61\pm 0.07$ 
&$5.21^{+8.00}_{-5.21}$, $0.371\pm 0.107$, $0.135\pm 0.042$ 
\\	
$\Gamma_{\Lambda_c D_s},f_{\Lambda_c D_s}$
&11.0, 0.41$^*$
&---
&---
\\
${\cal B}_{J/\Lambda}$, ${\cal B}_{\eta_c\Lambda}$ 
&$(4.3,8.6)\times10^{-2}$
&$(70,30)\times10^{-2}$
&$(80,20)\times10^{-2}$
\\
\hline
\end{tabular}
}
\end{table}
%
In our numerical analysis, the Cabibbo-Kobayashi-Maskawa (CKM) matrix elements in the Wolfenstein parameterization
are given by
$(V_{cb},V_{cs})=(A\lambda^2,1-\lambda^2/2)$, where
$A=0.826^{+0.016}_{-0.015}$ and $\lambda=0.22501\pm 0.00068$~\cite{pdg}.
Utilizing ${\cal M}_{a1}$ in Eq.~(\ref{fig1a}) and the input from~\cite{Hsiao:2023mud},
${\cal B}(B^-\to \Xi_c^0\bar \Lambda_c^-)=(7.8^{+2.3}_{-2.0})\times 10^{-4}$,
which is consistent with the data,
we extract $F=G=(0.172\pm 0.024)$~GeV$^2$.
The parameters in ${\cal M}(B^-\to M_{c\bar c} K^-)$ 
and ${\cal M}(B^-\to D^0\bar D_s^-)$ can be found in Refs.~\cite{Becirevic:2013bsa,pdg,MS}:
\begin{eqnarray}\label{a12}
&&
(f_{J/\psi},f_{\eta_c},f_{D_s})=(418, 387,250)~\text{MeV}\,,\nonumber\\
&&
(F_{1,0}^K,\sigma_{11}^K,\sigma_{12}^K,\sigma_{01}^K,\sigma_{02}^K,M_A)
=(0.36,0.43,0,0.70,0.27,5.4~\text{GeV})\,,\nonumber\\
&&
(F_{1,0}^D,\sigma_{11}^D,\sigma_{12}^D,\sigma_{01}^D,\sigma_{02}^D,M_A)
=(0.67,0.57,0,0.78,0,6.4~\text{GeV})\,. 
\end{eqnarray}
We thus calculate ${\cal B}(B^-\to M_{c\bar c} K^-,D^0\bar D_s^-)$
with the inputs in Eq.~(\ref{a12}), and compare them to the data~\cite{pdg}:
${\cal B}(B^-\to J/\psi K^-)=(1.020\pm 0.019)\times 10^{-3}$, 
${\cal B}(B^-\to \eta_c K^-)=(1.10\pm 0.07)\times 10^{-3}$, and 
${\cal B}(B^-\to D^0\bar D_s^-)=(9.0\pm 0.9)\times 10^{-3}$.
This results in the determination of
$a_2^{J/\psi}=0.263\pm 0.003$, $a_2^{\eta_c}=0.320\pm 0.010$, 
and $a_1^D=0.848\pm 0.043$.
Since $a_2^{J/\psi(\eta_c)}$ is of ${\cal O}(0.2-0.3)$ and
$a_1^D$ is of ${\cal O}(1.0)$, which is in agreement with the empirical inputs 
in the generalized factorization~\cite{Bauer:1986bm,Ali:1998eb,
Hsiao:2019wyd,Hsiao:2019ann,Hsiao:2020gtc,Hsiao:2021mlp,Hsiao:2022tfj,Geng:2005fh},
the determination is demonstrated to be applicable. 

We obtain $g_{pD}=13.2$ from~\cite{Duan:2023dky} 
and $g_{pK}=-11.3$ from~\cite{Yang:2018idi}.
The coupling constants $f_{\Xi_c D}$, $f_{\Lambda\eta_c}$, 
$(g_{\Lambda J},h_{\Lambda J})$, and
$f_{\Lambda_c D_s}$ are associated with the decays of
${\cal P}_{cs}^0\to \Xi_c^0 \bar D^0$, 
${\cal P}_{cs}^0\to \eta_c\Lambda$,
${\cal P}_{cs}^0\to J/\psi\Lambda$, and
${\cal P}_{cs}^0\to \Lambda_c^+\bar D_s^-$, respectively.
The studies of the ${\cal P}_{cs}^0$ decays 
in~Refs.~\cite{Ortega:2022uyu}, \cite{Azizi:2023iym}, and \cite{Wang:2024rsm}
lead to three different sets of the coupling constants in 
the scenarios S1, S2, and S3. 

\begin{figure}[t!]
\includegraphics[width=2.5in]{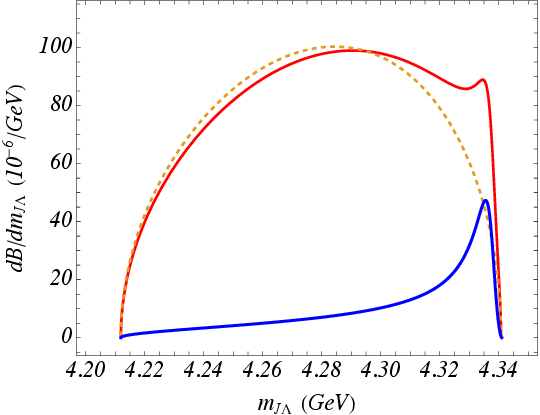}
\caption{Differential branching fractions of $B^-\to J/\psi\Lambda\bar p$
are shown as a function of $m_{J\Lambda}$, where
the blue-solid line represents the resonant ${\cal P}_{cs}$ state,
peaking around 4.338~GeV, while the golden-yellow dashed line 
corresponds to the non-resonant contribution, 
as studied in Refs.~\cite{Hsiao:2014tda,Hsiao:2013dta,Cheng:2012fq,Chen:2008sw}.
Additionally, the red-solid line accounts for the combined effects of 
the resonant and non-resonant contributions,
along with the destructive interference.}\label{fig2}
\end{figure}

Utilizing the parameter inputs for
the rescattering amplitudes in Eqs.~(\ref{fig1a}), (\ref{fig1b}), (\ref{fig1c}), and (\ref{amp_abc}),
we calculate the rescattering branching fractions. Nonetheless,
the cutoff parameter varies between different decays
and is poorly known phenomenologically~\cite{Tornqvist:1993ng}.
Since the rescattering amplitude of Fig.~\ref{fig1}(c) suggests that 
the cutoff parameter $\Lambda_{\rm cut}$ should be no less than
the mass of the exchange particle $\bar\Lambda_c^-$~\cite{Cheng:2004ru},
we set $\Lambda_{\rm cut}=(3\pm 1)$~GeV,
with $\delta \Lambda_{\rm cut}=1$~GeV to illustrate the sensitivity.
Our results are presented in Table~\ref{tab2}, particularly
${\cal B}(B^-\to \bar p({\cal P}_{cs}^0\to)J/\psi\Lambda,\bar p({\cal P}_{cs}^0\to)\eta_c\Lambda)$.
Additionally, we depict the $m_{J\Lambda}$ invariant spectrum of $B^-\to J/\psi\Lambda\bar p$
in Fig.~\ref{fig2}, which is based on the equations in Eq.~(\ref{3b_amp}) and Eq.~(\ref{gamma1}).
%
%
%
\begin{table}[b!]
\caption{Rescattering branching fractions calculated in the scenarios S1, S2, and S3. 
The first error incorporates uncertainties from the weak and strong couplings in the triangle loop, 
and the second error comes from $\delta \Lambda_{\rm cut}$.  
Total branching fraction ${\cal B}(B^-\to {\cal P}_{cs}^0\bar p)$ is the sum of 
these rescattering branching fractions. 
Branching fractions of the resonant three-body decays
$B^-\to\bar p({\cal P}_{cs}^0\to)J/\psi\Lambda$ and
$B^-\to\bar p({\cal P}_{cs}^0\to)\eta_c\Lambda$ are obtained with the approximate relation: 
${\cal B}(B^-\to {\cal P}_{cs}^0\bar p,{\cal P}_{cs}^0\to M_{c\bar c}\Lambda) 
\simeq {\cal B}(B^-\to {\cal P}_{cs}^0\bar p)\times  {\cal B}({\cal P}_{cs}^0\to M_{c\bar c} \Lambda)$, 
where the error combines the possible uncertainties from the two branching fractions.
}\label{tab2}
{\tiny
\begin{tabular}{|l|c|c|c|} 
\hline
&S1 (Ortega)~\cite{Ortega:2022uyu}
&S2 (Azizi)~\cite{Azizi:2023iym}
&S3 (Wang)~\cite{Wang:2024rsm}\\
\hline
${\cal B}(B^-\to \Xi_c^0\bar \Lambda_c^-\to {\cal P}_{cs}^0\bar p)$
&$(2.9\pm 0.8^{+3.0}_{-2.3})\times 10^{-8}$
&---
&---
\\
${\cal B}(B^-\to \eta_c K^-\to {\cal P}_{cs}^0\bar p)$
&$(5.2\pm 0.3^{+1.1}_{-2.2})\times 10^{-9}$
&$(2.8^{+1.1+0.6}_{-0.9-1.2})\times 10^{-9}$
&$(2.7^{+1.6+0.6}_{-1.2-1.2}) \times 10^{-9}$
\\	
${\cal B}(B^-\to J/\psi K^-\to {\cal P}_{cs}^0\bar p)$
&$(5.9\pm 0.1^{+64.6}_{-\;\,4.1})\times 10^{-7}$
&$(1.5\pm 0.3^{+19.0}_{-\;\,1.0})\times 10^{-7}$
&$(1.7^{+1.2+2.3}_{-0.9-1.1})\times 10^{-6}$
\\	
${\cal B}(B^-\to D^0\bar D_s^-\to{\cal P}_{cs}^0\bar p)$
&$(6.9\pm 0.7^{+3.6}_{-6.7})\times 10^{-6}$
&---
&---
\\
\hline
${\cal B}(B^-\to {\cal P}_{cs}^0\bar p)$
&$(7.5\pm 0.7^{+7.4}_{-6.7})\times10^{-6}$
&$(1.5\pm 0.3^{+19.0}_{-\;\,1.0})\times10^{-7}$
&$(1.7^{+1.2+2.3}_{-0.9-1.1})\times10^{-6}$
\\
${\cal B}(B^-\to {\cal P}_{cs}^0\bar p,{\cal P}_{cs}^0\to J/\psi\Lambda)$
&$(3.2^{+3.2}_{-2.9})\times10^{-7}$
&$(1.1^{+13.1}_{-\;\,0.7})\times10^{-7}$
&$(1.4^{+2.1}_{-1.1})\times10^{-6}$
\\
${\cal B}(B^-\to {\cal P}_{cs}^0\bar p,{\cal P}_{cs}^0\to\eta_c\Lambda)$
&$(6.5^{+6.4}_{-5.8})\times10^{-7}$
&$(0.5^{+5.9}_{-0.3})\times10^{-7}$
&$(3.2^{+5.2}_{-2.8})\times10^{-7}$
\\
\hline
\end{tabular}
}
\end{table}
%

\section{Discussions and Conclusion}
The resonant $B^-\to {\cal P}_{cs}^0 \bar p,{\cal P}_{cs}^0\to J/\psi \Lambda$ decay
relies on the final state interaction. In our study, 
we consider potential triangle rescattering processes, 
as illustrated in Fig.~\ref{fig1}(a, b, c). Our calculation requires
the coupling constants from ${\cal P}_{cs}^0$ strong decays.
Information from Refs.~\cite{Ortega:2022uyu}, \cite{Azizi:2023iym}, and \cite{Wang:2024rsm}
results in the scenarios S1, S2, S3 for the coupling constants,
which are applied to our investigation. 

We hence obtain the numerical results for the resonant branching fractions 
listed in Table~\ref{tab2}, where the first and second error bars correspond to 
the uncertainties from the coupling constants and the cut-off parameter, respectively.
Additionally, it is found that some of these results are sensitive to $\delta \Lambda_{\rm cut}$, 
with $\delta \Lambda_{\rm cut}\simeq 30\%$ of $\Lambda_{\rm cut}$.

In S1~(Ortega)~\cite{Ortega:2022uyu},
${\cal P}_{cs}^0$ is studied to consist of 45\% of $\Lambda_c\bar D_s^-$, 
28\% of $\Lambda_c\bar D_s^{*-}$, and other less significant molecular components
in the constituent quark model.
Using $\Gamma_{\rm th}=\Gamma_{\Xi_c D}+\Gamma_{\eta_c\Lambda}
+\Gamma_{J\Lambda}+\Gamma_{\Lambda_c D_s}$,
the total decay width $\Gamma_{\rm th}=13.9$~MeV 
is twice as large as $\Gamma_{\rm ex}=(7.0\pm 1.2\pm 1.3)$~MeV,
due to ${\cal P}_{cs}^0$ significantly decaying into $\Lambda_c^+\bar D_s^-$.
We adopt the decay widths in~\cite{Ortega:2022uyu}, and 
extract the coupling constants listed in Table~\ref{tab1},
assuming $g_{J\Lambda}=h_{J\Lambda}$.
In this scenario, 
${\cal B}(B^-\to D^0\bar D_s^-\to{\cal P}_{cs}^0\bar p)
=(6.9\pm 0.7^{+3.6}_{-6.7})\times 10^{-6}$ is calculated 
to be the dominated rescattering branching fraction.
It is found that ${\cal B}\sim10^{-5}$ is influenced by two enhancing factors:
the initial weak decay $B^-\to D^0\bar D_s^-$, of which 
the branching fraction is around 10 times larger than the other initial branching fractions,
and the significant decay width $\Gamma_{\Lambda_c D_s}$. 
However, ${\cal B}_{J\Lambda}=4.3\times10^{-2}$
results in ${\cal B}(B^-\to \bar p({\cal P}_{cs}^0\to)J/\psi\Lambda)
=(3.2^{+3.2}_{-2.9})\times10^{-7}$, which is six times smaller than 
${\cal B}_{\rm ex}(B^-\to \bar p({\cal P}_{cs}^0\to)J/\psi\Lambda)
=(1.8\pm 0.3)\times 10^{-6}$ in Eq.~(\ref{data1}).

In S2~(Azizi)~\cite{Azizi:2023iym}, 
${\cal P}_{cs}^0$ is considered to be a $\Xi_c\bar D$ molecule. 
Using QCD sum rules from~\cite{Azizi:2023iym},
${\cal P}_{cs}^0\to J/\psi\Lambda$ and ${\cal P}_{cs}^0\to \eta_c\Lambda$
are studied as the dominant decay channels. Additionally,
both the decay widths and coupling constants are calculated, leading to
$\Gamma_{\eta_c \Lambda}\simeq \Gamma_{J\Lambda}/2$ and
$\Gamma_{\rm th}\simeq \Gamma_{\eta_c \Lambda}+\Gamma_{J\Lambda}=(10.40\pm1.93)$~MeV, 
which is more consistent with $\Gamma_{\rm ex}$ than that in S1.
Consequently, we obtain ${\cal B}(B^-\to J/\psi K^-\to {\cal P}_{cs}^0\bar p)
=(1.5\pm 0.3^{+19.0}_{-\;\,1.0})\times 10^{-7}$ as the leading rescattering branching fraction. 
This results in ${\cal B}(B^-\to \bar p({\cal P}_{cs}^0\to)J/\psi\Lambda)
=(1.1^{+13.1}_{-\;\,0.7})\times10^{-7}$. 
While the central value is approximately sixteen times smaller than the data,
the error range leads to a slight possibility of reaching the experimental result.

In S3~(Wang)~\cite{Wang:2024rsm},
${\cal P}_{cs}^0$ is also identified as a $\Xi_c\bar D$ molecule. 
When QCD sum rules are applied~\cite{Wang:2024rsm,Wang:2022neq},
$\Xi_c$ is regarded as the $csq$ color-singlet cluster, whose 
quantum numbers are the same as those of the physical anti-triplet charm baryon.
Meanwhile, $\bar D$ is considered as the $c\bar q$ color-singlet cluster, 
sharing the same quantum numbers with the physical charmed meson. 
Consequently, ${\cal P}_{cs}^0\to J/\psi\Lambda,\,\eta_c\Lambda$ 
are calculated as the dominant decay channels, 
with both the decay widths and coupling constants provided.
Although $\Gamma_{J\Lambda}$ presents a large error in Table~\ref{tab1},
$\Gamma_{\rm th}\simeq \Gamma_{\eta_c \Lambda}+\Gamma_{J\Lambda}
=6.16$~MeV is close to $\Gamma_{\rm ex}=7.0$~MeV. However,
$g_{J\Lambda}$ and $h_{J\Lambda}$ do not deviate significantly from each other
as they do in S2, where $h_{J\Lambda}\gg g_{J\Lambda}\simeq 0$. 
Note that both S2 and S3 use the same model, but they lead to slightly different results.
We thus obtain ${\cal B}(B^-\to J/\psi K^-\to {\cal P}_{cs}^0\bar p)
=(1.7^{+1.2+2.3}_{-0.9-1.1})\times 10^{-6}$, and
the subsequent branching fraction ${\cal B}(B^-\to \bar p({\cal P}_{cs}^0\to)J/\psi\Lambda)
=(1.4^{+2.1}_{-1.1})\times10^{-6}$, which can interpret the data.

The pentaquark candidate ${\cal P}_{cs}^0$ has been extensively studied
in various models~\cite{Ortega:2022uyu,Azizi:2023iym,
Wang:2024rsm,Wang:2022neq,Yan:2022wuz,Zhu:2022wpi,
Meng:2022wgl,Ozdem:2022kei,Nakamura:2022gtu,Wang:2023ael,
Wang:2023eng,Giachino:2022pws,Chen:2022wkh}, and
the feasibility of these models depends on whether the predicted cousin pentaquarks
can be discovered in future measurements. In our new findings, 
as long as the ${\cal P}_{cs}^0$ strong decays are studied
similar to~\cite{Ortega:2022uyu,Azizi:2023iym,Wang:2024rsm}, 
potential triangle rescattering effects can also be used to test models.
Since ${\cal B}(B^-\to \bar p({\cal P}_{cs}^0\to)J/\psi\Lambda)$ in S3
is calculated to agree with the data, rather than the other two scenarios,
the approach of QCD sum rules in~\cite{Wang:2024rsm}
appears to provide the most suitable information. 
On the other hand, 
we need to await the available coupling constants of the ${\cal P}_c^\pm$ decays 
to conduct a systematic analysis of 
$\bar B^0_s\to \bar p({\cal P}_c^+\to)J/\psi p+p({\cal P}_c^-\to) J/\psi \bar p$.

Since the invariant mass spectrum allows for a more detailed examination,
we depict the $m_{J\Lambda}$ invariant spectrum of $B^-\to J/\psi\Lambda\bar p$ in Fig.~\ref{fig2}.
Using the parameter inputs from~S3, 
the resonant contribution, shown as the blue-solid line, amounts to ${\cal B}_{\rm RE}\simeq 1.3\times 10^{-6}$, 
consistent with the value in Table~\ref{tab2}, where the approximation relation is applied. 
The non-resonant differential branching fraction, depicted as the golden-yellow dashed line,
serves as the background to the peak caused by the resonant ${\cal P}_{cs}$ state,
integrating to ${\cal B}_{\rm NR}\simeq 10.0\times 10^{-6}$. 

Specifically,
in the (4.32-4.34)~GeV resonant region, the non-resonant contribution leads to
${\cal B}\sim 1.1\times 10^{-6}$, comparable to the pentaquark contribution.
Interestingly, it is found that the resonant and non-resonant amplitudes interfere destructively, 
leading to $|{\cal B}_{\rm INF}|\simeq 0.9\times 10^{-6}$. As a result, the total branching fraction, 
${\cal B}_{\rm T}={\cal B}_{\rm RE}+{\cal B}_{\rm NR}+{\cal B}_{\rm INF}\simeq 10.4\times 10^{-6}$,
is consistent with the experimental data~\cite{pdg}. Incorporating the RE, NR, and INF
contributions results in the red-solid line in Fig.~\ref{fig2}, 
where the resonant peak remains clearly visible.

In conclusion, we have considered that 
the final state interaction (FSI) is responsible for the formation of
${\cal P}_{cs}^0$ in the resonant three-body decay
$B^-\to {\cal P}_{cs}^0\bar p$, ${\cal P}_{cs}^0\to J/\psi \Lambda$.
By interpreting ${\cal P}_{cs}^0$ as a $\Xi_c\bar D$ molecule,
decaying into $J/\psi\Lambda$ and $\eta_c\Lambda$, 
we have identified the dominant FSI triangle rescattering effect,
initiated by the weak decay $B^-\to J/\psi K^-$. 
With the $\bar\Lambda$ anti-baryon exchange, 
$J/\psi$ and $K^-$ proceed to rescatter,
transforming into ${\cal P}_{cs}^0$ and $\bar p$, respectively.
We have thus calculated ${\cal B}(B^-\to \bar p({\cal P}_{cs}^0\to)J/\psi\Lambda)
=(1.4^{+2.1}_{-1.1})\times10^{-6}$ to interpret the data.

\section*{ACKNOWLEDGMENTS}
We would like to thank Dr.~Qi Wu for useful discussions.
This work was supported in part
by National Science Foundation of China (Grants No.~12175128 and No.~11675030)
and Innovation Project of Graduate Education in Shanxi Province (2023KY430).



\begin{thebibliography}{99}
\bibitem{Chen:2022asf}
H.~X.~Chen, W.~Chen, X.~Liu, Y.~R.~Liu and S.~L.~Zhu,
Rept. Prog. Phys. \textbf{86}, 026201 (2023). 

\bibitem{Johnson:2024omq}
D.~Johnson, I.~Polyakov, T.~Skwarnicki and M.~Wang,
arXiv:2403.04051 [hep-ex].

\bibitem{LHCb:2015yax}
R.~Aaij \textit{et al.} [LHCb],
Phys. Rev. Lett. \textbf{115}, 072001 (2015). 

\bibitem{LHCb:2019kea}
R.~Aaij \textit{et al.} [LHCb],
Phys. Rev. Lett. \textbf{122}, 222001 (2019). 

\bibitem{LHCb:2020jpq}
R.~Aaij \textit{et al.} [LHCb],
Sci. Bull. \textbf{66}, 1278 (2021). 

\bibitem{Brodsky:1997yr}
S.~J.~Brodsky and F.~S.~Navarra,
Phys. Lett. B \textbf{411}, 152 (1997). 

\bibitem{LHCb:2021chn}
R.~Aaij \textit{et al.} [LHCb],
Phys. Rev. Lett. \textbf{128}, 062001 (2022). 

\bibitem{LHCb:2022ogu}
R.~Aaij \textit{et al.} [LHCb],
Phys. Rev. Lett. \textbf{131}, 031901 (2023). 

\bibitem{pdg}
S. Navas \textit{et al.} (Particle Data Group), Phys. Rev. D \textbf{110}, 030001 (2024).

\bibitem{Nakamura:2021dix}
S.~X.~Nakamura, A.~Hosaka and Y.~Yamaguchi,
Phys. Rev. D \textbf{104}, L091503 (2021). 

\bibitem{Yan:2021nio}
M.~J.~Yan, F.~Z.~Peng, M.~S\'anchez S\'anchez and M.~Pavon Valderrama,
Eur. Phys. J. C \textbf{82}, 574 (2022). 

\bibitem{Wu:2024lud}
Q.~Wu and D.~Y.~Chen,
Phys. Rev. D \textbf{109}, 094003 (2024). 

\bibitem{Chen:2008sw}
C.~H.~Chen, H.~Y.~Cheng, C.~Q.~Geng and Y.~K.~Hsiao,
Phys. Rev. D \textbf{78}, 054016 (2008). 

\bibitem{Hsiao:2009mk}
Y.~K.~Hsiao, 
Int. J. Mod. Phys. A \textbf{24}, 3638 (2009). 

\bibitem{Hou:2000bz}
W.~S.~Hou and A.~Soni, Phys.\ Rev.\ Lett.\  {\bf 86}, 4247 (2001).

\bibitem{Suzuki:2006nn}
M.~Suzuki, 
J. Phys. G \textbf{34}, 283 (2007). 

\bibitem{Hsiao:2014zza}
Y.~K.~Hsiao and C.~Q.~Geng,
Phys. Rev. D \textbf{91}, 077501 (2015). 

\bibitem{Hou:2019uxa}
W.~S.~Hou, M.~Kohda, T.~Modak and G.~G.~Wong,
Phys.\ Lett.\ B {\bf 800}, 135105 (2020). 

\bibitem{Hsiao:2019wyd}
Y.~K.~Hsiao, S.~Y.~Tsai, C.~C.~Lih and E.~Rodrigues,
JHEP \textbf{04}, 035 (2020). 

\bibitem{Chua:2022wmr}
C.~K.~Chua,
Phys. Rev. D \textbf{106}, 036015 (2022). 

\bibitem{Huang:2021qld}
X.~Huang, Y.~K.~Hsiao, J.~Wang and L.~Sun, 
Adv. High Energy Phys. \textbf{2022}, 4343824 (2022). 

\bibitem{Hsiao:2023mud}
Y.~K.~Hsiao, 
JHEP \textbf{11}, 117 (2023). 

\bibitem{LHCb:2022oyl}
R.~Aaij \textit{et al.} [LHCb],
Phys. Rev. D \textbf{108}, 012007 (2023). 

\bibitem{Brodsky:1980pb}
S.~J.~Brodsky, P.~Hoyer, C.~Peterson and N.~Sakai, 
Phys. Lett. B \textbf{93}, 451 (1980).

\bibitem{Brodsky:1981se}
S.~J.~Brodsky, C.~Peterson and N.~Sakai, 
Phys. Rev. D \textbf{23}, 2745 (1981).

\bibitem{Brodsky:1997fj} 
S.~J.~Brodsky and M.~Karliner, Phys.\ Rev.\ Lett.\  {\bf 78}, 4682 (1997).

\bibitem{Mikhasenko:2012km} 
M.~Mikhasenko, 
Phys.\ Atom.\ Nucl.\  {\bf 77}, 623 (2014) [Yad.\ Fiz.\  {\bf 77}, 658 (2014)].

\bibitem{Chang:2001iy} 
C.~H.~V.~Chang and W.~S.~Hou, Phys.\ Rev.\ D {\bf 64}, 071501 (2001).

\bibitem{Brodsky:2001yt} 
S.~J.~Brodsky and S.~Gardner, Phys.\ Rev.\ D {\bf 65}, 054016 (2002).

\bibitem{Zhou:2017bhq}
P.~Zhou, Y.~K.~Hsiao and C.~Q.~Geng, 
Annals Phys. \textbf{383}, 278 (2017). 

\bibitem{Hsiao:2015nna}
Y.~K.~Hsiao and C.~Q.~Geng,
Phys. Lett. B \textbf{751}, 572 (2015). 

\bibitem{Ortega:2022uyu}
P.~G.~Ortega, D.~R.~Entem and F.~Fernandez,
Phys. Lett. B \textbf{838}, 137747 (2023). 

\bibitem{Azizi:2023iym}
K.~Azizi, Y.~Sarac and H.~Sundu,
Phys. Rev. D \textbf{108}, 074010 (2023). 

\bibitem{Wang:2024rsm}
X.~W.~Wang and Z.~G.~Wang,
Phys. Rev. D \textbf{110}, 014008 (2024). 

\bibitem{Burns:2022uha}
T.~J.~Burns and E.~S.~Swanson,
Phys. Lett. B \textbf{838}, 137715 (2023). 

\bibitem{Duan:2023dky}
M.~X.~Duan, L.~Qiu, X.~Z.~Ling and Q.~Zhao,
Phys. Rev. D \textbf{109}, L031507 (2024). 

\bibitem{Gourdin:1994xx}
M.~Gourdin, Y.~Y.~Keum and X.~Y.~Pham,
Phys. Rev. D \textbf{51}, 3510 (1995). 

\bibitem{Cheng:1998kd}
H.~Y.~Cheng and K.~C.~Yang,
Phys. Rev. D \textbf{59}, 092004 (1999). 

\bibitem{Yang:2018idi}
G.~S.~Yang and H.~C.~Kim,
Phys. Lett. B \textbf{785}, 434 (2018). 

\bibitem{MS} 
D.~Melikhov and B.~Stech, Phys.\ Rev.\ D {\bf 62}, 014006 (2000).

\bibitem{Hsiao:2019ait}
Y.~K.~Hsiao, Y.~Yu and B.~C.~Ke,
Eur. Phys. J. C \textbf{80}, 895 (2020). 

\bibitem{Yu:2020vlt}
Y.~Yu and Y.~K.~Hsiao,
Phys. Lett. B \textbf{820}, 136586 (2021). 

\bibitem{Yu:2021euw}
Y.~Yu, Y.~K.~Hsiao and B.~C.~Ke,
Eur. Phys. J. C \textbf{81}, 1093 (2021). 

\bibitem{Hsiao:2021tyq}
Y.~K.~Hsiao and Y.~Yu,
Phys. Rev. D \textbf{104}, 034008 (2021). 

\bibitem{Du:2021zdg}
M.~C.~Du and Q.~Zhao,
Phys. Rev. D \textbf{104}, 036008 (2021). 

\bibitem{Tsai:2021ota}
S.~Y.~Tsai and Y.~K.~Hsiao,
arXiv:2107.03634 [hep-ph].

\bibitem{Hsiao:2023qtk}
Y.~K.~Hsiao, S.~Q.~Yang, W.~J.~Wei and B.~C.~Ke,
JHEP \textbf{12}, 226 (2025). 

\bibitem{Hsiao:2014tda}
Y.~K.~Hsiao and C.~Q.~Geng, 
Eur. Phys. J. C \textbf{75}, 101 (2015). 

\bibitem{Hsiao:2013dta}
Y.~K.~Hsiao and C.~Q.~Geng,
Phys. Lett. B \textbf{727}, 168 (2013). 

\bibitem{Cheng:2012fq}
H.~Y.~Cheng, C.~Q.~Geng and Y.~K.~Hsiao,
Phys. Rev. D \textbf{89}, 034005 (2014). 

\bibitem{Geng:2005fh}
C.~Q.~Geng and Y.~K.~Hsiao,
Phys. Lett. B \textbf{619}, 305 (2005). 

\bibitem{Geng:2007cw}
C.~Q.~Geng, Y.~K.~Hsiao and J.~N.~Ng,
Phys. Rev. D \textbf{75}, 094013 (2007). 
\bibitem{Becirevic:2013bsa}
D.~Be\v{c}irevi\'c, G.~Duplan\v{c}i\'c, B.~Klajn, B.~Meli\'c and F.~Sanfilippo,
Nucl.~Phys.~B~\textbf{883}, 306 (2014). 



\bibitem{Bauer:1986bm}
M.~Bauer, B.~Stech and M.~Wirbel,
Z. Phys. C \textbf{34}, 103 (1987).

\bibitem{Ali:1998eb}
A.~Ali, G.~Kramer and C.~D.~Lu,
Phys. Rev. D \textbf{58}, 094009 (1998). 

\bibitem{Hsiao:2019ann}
Y.~K.~Hsiao, S.~Y.~Tsai and E.~Rodrigues,
Eur. Phys. J. C \textbf{80}, 565 (2020). 

\bibitem{Hsiao:2020gtc}
Y.~K.~Hsiao, L.~Yang, C.~C.~Lih and S.~Y.~Tsai,
Eur. Phys. J. C \textbf{80}, 1066 (2020). 

\bibitem{Hsiao:2021mlp}
Y.~K.~Hsiao and C.~C.~Lih,
Phys. Rev. D \textbf{105}, 056015 (2022). 

\bibitem{Hsiao:2022tfj}
Y.~K.~Hsiao, 
Phys. Lett. B \textbf{845}, 138158 (2023). 

\bibitem{Tornqvist:1993ng}
N.~A.~Tornqvist,
Z. Phys. C \textbf{61}, 525 (1994). 

\bibitem{Cheng:2004ru}
H.~Y.~Cheng, C.~K.~Chua and A.~Soni, 
Phys. Rev. D \textbf{71}, 014030 (2005). 

\bibitem{Wang:2022neq}
X.~W.~Wang and Z.~G.~Wang,
Chin. Phys. C \textbf{47}, 013109 (2023). 

\bibitem{Yan:2022wuz}
M.~J.~Yan, F.~Z.~Peng, M.~S\'anchez S\'anchez and M.~Pavon Valderrama,
Phys. Rev. D \textbf{107}, 074025 (2023). 

\bibitem{Zhu:2022wpi}
J.~T.~Zhu, S.~Y.~Kong and J.~He,
Phys. Rev. D \textbf{107}, 034029 (2023). 

\bibitem{Meng:2022wgl}
L.~Meng, B.~Wang and S.~L.~Zhu,
Phys. Rev. D \textbf{107}, 014005 (2023). 

\bibitem{Ozdem:2022kei}
U.~\"Ozdem, 
Phys. Lett. B \textbf{836}, 137635 (2023). 

\bibitem{Nakamura:2022gtu}
S.~X.~Nakamura and J.~J.~Wu, 
Phys. Rev. D \textbf{108}, L011501 (2023). 

\bibitem{Wang:2023ael}
F.~L.~Wang and X.~Liu,
Phys. Rev. D \textbf{109}, 1 (2024). 

\bibitem{Wang:2023eng}
B.~Wang, K.~Chen, L.~Meng and S.~L.~Zhu, 
Phys. Rev. D \textbf{109}, 074035 (2024). 

\bibitem{Giachino:2022pws}
A.~Giachino, A.~Hosaka, E.~Santopinto, S.~Takeuchi, M.~Takizawa and Y.~Yamaguchi,
Phys. Rev. D \textbf{108}, 074012 (2023). 

\bibitem{Chen:2022wkh}
K.~Chen, Z.~Y.~Lin and S.~L.~Zhu,
Phys. Rev. D \textbf{106}, 116017 (2022). 

\end{thebibliography}
\end{document}